\documentclass[english,aip,pop,reprint,floatfix]{revtex4-2}
\usepackage[T1]{fontenc}
\usepackage[latin9]{inputenc}
\usepackage{amsmath}
\usepackage{graphicx}
\usepackage[normalem]{ulem} 
\usepackage{babel}
\usepackage{color}
\usepackage{url}
\usepackage{subcaption}
\usepackage{array}
\newcolumntype{L}{>{$}l<{$}}
\begin{document}
\title{Forced Reconnection in Voigt-Regularized MHD}

\author{Andrew Brown}
\affiliation{Department of Astrophysical Sciences, Princeton University, New Jersey 08543, USA}
\email{aobrown@princeton.edu}
\author{Yi-Min Huang}
\affiliation{Department of Astronomy, University of Maryland, College Park, MD
20742}
\affiliation{National Aeronautics and Space Administration Goddard Space Flight
Center, Greenbelt, MD 20771}
\author{Amitava Bhattacharjee}
\affiliation{Department of Astrophysical Sciences, Princeton University, New Jersey
08543, USA}
\begin{abstract}
Forced reconnection in Voigt-regularized MHD is investigated in the Hahm-Kulsrud-Taylor problem. It is shown that Voigt regularization introduces an early linear phase of reconnection that partially bypasses the ideal current sheet formation phase. Three distinct regimes of Rutherford-like island growth and saturation are elucidated, including the braking effects of regularization and viscosity. It is conjectured, with numerical justification, that the inclusion of drag in the momentum equation does not inhibit the relaxation to steady state in the long-time limit. 
\end{abstract}
\maketitle

\section{Introduction}
The design of stellarators -- toroidal magnetic confinement devices with shaped, three-dimensional (3D) magnetic fields -- requires precise computation of magnetohydrostatic (MHS) equilibria, or magnetic fields $\boldsymbol{B}$ and scalar pressure $p$ satisfying
\begin{equation}
    \boldsymbol{J}\times \boldsymbol{B} = \nabla p,
\end{equation}
where the current $\boldsymbol{J} = \nabla \times \boldsymbol{B}$. The large design space for stellarator optimization necessitates that any equiliibrium computation be rapid and iterable\citep{imbert2025}.

In the axisymmetric case (\textit{viz}., tokamak equilibria), the toroidal angle acts as a cyclic coordinate in the magnetic field Hamiltonian, guaranteeing nested flux surfaces in perfect axisymmetry. The MHS condition reduces to the Grad-Shafranov equation, an elliptic PDE for a scalar flux function\citep{Freidberg1987}. When this symmetry is broken, nested flux surfaces break into islands and, potentially, regions of stochastic field lines. It is an open question, in fact, whether 3D magnetic fields with nested flux surfaces and smooth pressure profile exist at all\citep{grad1967}. 

Nevertheless, many extant equilibrium (e.g., VMEC and DESC) codes assume perfectly nested flux surfaces throughout the plasma volume\citep{hirshmanw1983, dudtk2020}. State-of-the-art stellarator optimization relies heavily on these codes \citep{imbert2025}. Several equilibrium codes (e.g., SIESTA, HINT, and SPEC) which do not assume nested flux surfaces have been developed\citep{HirshmanSC2011, Suzuki2017, hudsonddhmnl2012}, each relying on their own scheme to find MHS equilibria, but none of them provably, stably converge to smooth MHS equilibria. 

Voigt regularization first appeared in Oskolkov's treatment of visco-elastic fluids\citep{oskolkov1973}. Oskolkov found that a Kelvin-Voigt constitutive relation introduced a Helmholtz operator acting on the time derivatives in the fluid equations of motion. Such operators were later investigated as regularization schemes in fluid mechanics\citep{cao2006}. In recent work\citep{constantinp2023}, it was shown that a Voigt-regularized version of incompressible MHD will provably converge to an equilibrium that is not a force-free Taylor state\citep{taylor1974}, but retains a nontrivial pressure gradient.

In the first numerical investigation of Voigt-MHD\citep{huang2025}, it was found that 2D relaxation is frustrated by small-scale structures that are slow to relax. However, the addition of resistivity and a compensatory external electric field was sufficient to drive the system to equilibrium. Regularization was found to slow dynamics, but the CFL stability condition\citep{presstvf2002} on explicit timestep size was significantly relaxed. Heuristically speaking, in wave-vector ($\boldsymbol{k}$) space, the regularization acts on all eigenfrequencies by $\omega \to \omega (1 + \alpha k^2)^{-1}$, so the maximum stable timestep in the unregularized case $\omega(k_{max})^{-1}$ is replaced by a larger value $\sim \omega(\alpha^{-1/2})^{-1}$. The net effect was, in the best-performing case, approximately two orders of magnitude faster convergence. However, the addition of resistivity produced $O(\eta)$ residual flow. Though it is true that real devices have small flows in equilibrium, this deviation from MHS force balance is an undesirable complication to the 3D equilibrium problem -- itself already dauntingly complex. We demonstrate here that if we generalize our earlier Voigt model \citep{huang2025} by including resistivity as well as friction, the small flows are eliminated and the system relaxes asymptotically to MHS equilibria.

The following is the plan of this paper. We begin by elucidating the preliminary findings of our previous work\citep{huang2025}. In section \ref{sec:voigt}, we present the Voigt-MHD system and the Hahm-Kulsrud-Taylor (HKT) forced reconnection test problem that we will analyze throughout the work. In section \ref{sec:lin}, we present an early-time, linear theory of Voigt reconnection, demonstrating that reconnection begins well before an ideal current sheet can form. In section \ref{sec:nonlin}, we modify the classical Rutherford model for island evolution to account for the effects of viscosity, friction, and Voigt regularization. Finally, in section \ref{sec:equil} we present numerical evidence for our conjecture that the Constantin-Pasqualotto theorem generalizes to the case considered here, which includes resistivity, a driven external field, and friction.

\section{Voigt-Regularized MHD}
\label{sec:voigt}
We study the incompressible Voigt-regularized MHD equations in the form
\begin{align}
\partial_{t}&\left(\boldsymbol{u}-\alpha_{1}\nabla^{2}\boldsymbol{u}\right)+\boldsymbol{u}\cdot\nabla\boldsymbol{u}=\nonumber\\
&-\nabla p+\boldsymbol{J}\times\boldsymbol{B}+\nu\nabla^{2}\boldsymbol{u} - \mu \boldsymbol{u},\label{eq:momentum}
\end{align}
\begin{equation}
\partial_{t}\left(\boldsymbol{B}-\alpha_{2}\nabla^{2}\boldsymbol{B}\right)=\nabla\times\left(\boldsymbol{u}\times\boldsymbol{B}-\eta\boldsymbol{J}-\boldsymbol{E}_{\text{ext}}\right),\label{eq:induction}
\end{equation}
\begin{equation}
\boldsymbol{J}=\nabla\times\boldsymbol{B},\label{eq:J_from_B}
\end{equation}
\begin{equation}
\nabla\cdot\boldsymbol{u}=\nabla\cdot\boldsymbol{B}=0.\label{eq:constraints}
\end{equation}
When the regularization parameters $\alpha_1$ and $\alpha_2$ vanish, we recover the incompressible, visco-resistive MHD equations with uniform mass density, magnetic field $\boldsymbol{B}$, current $\boldsymbol{J}$, fluid velocity $\boldsymbol{u}$, resistivity $\eta$, viscosity $\nu$, friction coefficient $\mu$, and applied electric field $\boldsymbol{E_{ext}}$. These equations are identical to the form used in a recent investigation of Voigt-regularized MHD \citep{huang2025} except for the addition of a drag term $-\mu \boldsymbol{u}$ in the momentum equation. The regularizing term in the induction equation is similar to the effect of electron inertia\citep{grasso1999}, but the infinite-time current singularities of inertial MHD are avoided in Voigt MHD. See Appendix \ref{app:helicity} for details.

We now restrict our attention to our chosen test case, the classic Hahm-Kulsrud-Taylor problem \citep{hahmk1985}, anticipating that the insights gained from this model may be generalized in future work. We work in a 2D slab $x\in [-L_x/2,L_x/2]$ and $y\in [-L_y/2,L_y/2)$ with conducting wall boundary conditions in $x$, periodic boundary conditions in $y$, and symmetry in the $z$ direction.

We define the stream function $\phi$ and flux function $\psi$ such that
\begin{equation}
\begin{aligned}
    \boldsymbol{u} &= \hat{\boldsymbol{z}}\times\nabla\phi, \\\boldsymbol{B} &= \hat{\boldsymbol{z}}\times\nabla\psi + B_T \hat{\boldsymbol{z},}
    \label{eq:stream}
\end{aligned}
\end{equation}
and the induction equation takes the form\footnote{Here we correct a typographical error made in \cite{huang2025}.}
\begin{equation}
\partial_{t}\left(\psi-\alpha_{2}\nabla^{2}\psi\right)=\hat{\boldsymbol{z}}\cdot\boldsymbol{u}\times\boldsymbol{B}+\eta J-E_{\text{ext}}.\label{eq:induction2d}
\end{equation}
The momentum equation is
\begin{align}
    \partial_{t}\left(\boldsymbol{u}-\alpha_{1}\nabla^{2}\boldsymbol{u}\right)&+\boldsymbol{u}\cdot\nabla\boldsymbol{u}
    = \nonumber\\
    -\nabla p-JB_{y}\boldsymbol{\hat{x}}&+JB_{x}\boldsymbol{\hat{y}}+\nu\nabla^{2}\boldsymbol{u} - \mu \boldsymbol{u}.\label{eq:momentum2d}
\end{align}

The initial equilibrium has $\boldsymbol{u} =0$ and $\psi_0 = \frac{B_0 x^2}{2 a}$, which we perturb by $\psi_0\to\pm (a - \delta \cos ky)\psi_0$. Our numerical implementation takes $a=1.0$, $B_0 = 1.0$ for simplicity, and $L_x = 1.0$, $L_y = 7.0$ unless otherwise stated. The external field $E_{ext} =\eta J_0$, current $J = \nabla^2 \psi$, and the pressure evolution is determined self-consistently by enforcing incompressibility $\nabla\cdot\boldsymbol{u}$, so $p$ satisfies a Poisson equation
\begin{equation}
    \nabla^2 p = \nabla \cdot\left(-\boldsymbol{u}\cdot\nabla\boldsymbol{u}+\boldsymbol{J}\times\boldsymbol{B} \right).
    \label{eq:poisson}
\end{equation}

The system of equations \eqref{eq:stream}-\eqref{eq:poisson} is numerically implemented in a rectangular
domain with a Fourier-Chebyshev pseudospectral method using the Dedalus framework (\url{https://dedalus-project.org/}). The periodic ($y$) direction is represented by a Fourier representation truncated at $N_y = 128$ modes. The $x$ direction is discretized with a composite of multiple Chebyshev segments. We use three segments along the $x$ direction, with each segment employing $N_{x}=128$ Chebyshev polynomials. The middle segment is narrower to provide a higher resolution near the mid-plane. Specifically, the domain along the $x$ direction is within the region $\left[-L_{x}/2,L_{x}/2\right]$ and the middle segment corresponds to the region $\left[-L_{x}/20,L_{x}/20\right]$. A dealiasing factor of $3/2$ is applied in both directions.In this study, we use the third-order, four-stage ``RK443'' scheme.\citep{AscherRS1997} 

\section{Linear theory}
\label{sec:lin}
In the early linear phase, our treatment closely follows that of Hahm and Kulsrud. For simplicity, we will take the non-dissipative limit of the Voigt equations, so $\eta, \mu,\nu \to 0$, in this initial consideration.

The form of the perturbation forces the form
\begin{equation}
    \psi_1(x,y) = \psi_1(x) \cos (ky),
\end{equation}
and
\begin{equation}
    \phi(x, y) = \phi(x) \sin (ky).
\end{equation}
By symmetry, it suffices to consider the domain $x\geq 0 $ and require that $\psi_1(x)$ be even. For convenience, we define $\zeta = kB_0/a$. Factoring out $y$-dependence and assuming $\partial_x \gg k$, the linearized system becomes
\begin{equation}
\begin{aligned}
    \partial_t\left(1 - \alpha_2\partial_x^2\right)\psi_1 &= \zeta x \phi, \\
    \partial_t\left(1 - \alpha_1 \partial_x^2\right)\partial_x^2\phi &= - \zeta x \partial_x^2 \psi_1.
    \label{eq:HKT-linear}
\end{aligned}
\end{equation}

In the ideal case, we neglect $\alpha_1$ and $\alpha_2$ and define $\xi \equiv \psi_1/x$ to eliminate $\phi$, which yields the ideal equation

\begin{equation}
    \partial_t^2 \partial_x^2 \xi = -\zeta^2 x \partial_x^2(x \xi),
\end{equation}

which is manifestly invariant under the group of transformations parameterized by $\beta > 0$:
\begin{equation}
\begin{aligned}
    t &\mapsto \beta t,\\
    x &\mapsto \beta^{-1}x,
\end{aligned}
\end{equation}
so admits the well-known self-similarity solution $\xi(x,t) = \xi(x t)$,

\begin{align}
    \xi(xt) &= \frac{2}{\pi}\frac{a \zeta \delta}{\sinh ka}\int_{0}^{\zeta x t}\frac{\sin u}{u}du \\
    &=\frac{2}{\pi}\frac{a \zeta \delta}{\sinh ka} \text{Si}\left(\zeta x t\right).
\end{align}

The current channel lies in a thin layer around $x=0$ and shrinks $\sim1/t$. Just as resistive effects ``turn on" when gradients become sufficiently large, one expects that Voigt regularization will eventually become non-negligible when the current layer size is comparable to the Voigt scale $\sim\alpha^{1/2}$. We shall find, however, that Voigt regularization allows reconnection long before a singular current forms.

To get a rough sense of the effect of Voigt regularization on the ideal current sheet, we can integrate the induction equation in the form 
\begin{equation}
    \partial_t \psi_1(x, t) = \alpha_2 \partial_t \partial_x^2 \psi_{\text{ideal}}(x, t),
\end{equation}
substituting the self-similar $\psi_{\text{ideal}}$ on the RHS. This is analogous to the ``phase A" procedure employed by Hahm and Kulsrud.

We find
\begin{equation}
    \psi_1(x, t) \simeq \alpha_2 C \left[\zeta \cos \left(\zeta x t\right) + \frac{1}{x}\sin\left(\zeta x t\right)\right],
\end{equation}
so in the linear phase the growth is predicted to be linear in time,
\begin{equation}
    \psi_1(x=0,t) = \alpha_2 k^2\frac{4\delta B_0}{\pi \sinh (ka)}\frac{t}{\tau_A},
    \label{eq:naive-scaling}
\end{equation}
where $\tau_A = B_0/a$ is the Alfv\'{e}n time. The dimensionless factor $\alpha_2 k^2$ is a measure of the strength of the regularization, normalized by the perturbing mode length scale. It is natural to compare to the result of Hahm and Kulsrud, who found $\psi_1 \sim \frac{t^2}{\tau_R \tau_A}$  (where $\tau_R$ is the resistive timescale) in the linear phase.

This result is plausible, but a more detailed treatment reveals a subtlety that is not relevant in the ideal problem. To see this, we solve the initial value problem directly. The initial conditions are $\phi(x,t=0)=0$ and $\psi_1(x,t=0) = \delta B_0 \frac{x^2}{2 a}$, so we Laplace transform equation (\ref{eq:HKT-linear}), defining transformed functions $\psi_1(x,t)\xrightarrow{\mathcal{L}} \Psi(x,s)$ and $\phi(x,t) \xrightarrow{\mathcal{L}} \Phi(x,s)$ and recalling 
\begin{equation}
\mathcal{L}\{\partial_t\psi_1(x,t)\}= s \Psi - \psi_1(x,t)|_{t=0},
\end{equation}
we have the transformed system
\begin{equation}
\begin{aligned}
    s\left(1 - \alpha_2 \partial_x^2\right)\Psi - \delta B_0\frac{x^2}{2a} + \alpha_2\frac{\delta B_0}{a} &= \zeta x \Phi,\\
    s\left(1 - \alpha_1 \partial_x^2\right)\partial_x^2\Phi &= - \zeta x \partial_x^2 \Psi.
\end{aligned}
\label{eq:laplace}
\end{equation}

Once again we consider the ideal limit $\alpha_1 = \alpha_2 = 0$, so we solve
\begin{equation}
   \partial_x^2\Phi \left(1 + \frac{\zeta^2}{s^2}x^2\right) + 2 x \frac{\zeta^2}{s^2}\partial_x\Phi = - \frac{\zeta x}{s^2} \frac{\delta B_0}{a},
\end{equation}
and find
\begin{equation}
    \partial_x\Phi = \frac{C}{s^2 + \mu^2 x^2} - \frac{\zeta \delta B_0 x^2}{2 a \left(s^2 +\zeta^2 x^2\right)}.
\end{equation}
Integrating and setting the irrelevant constant to zero,
\begin{equation}
    \Phi = -\frac{\delta B_0}{2 a \zeta}x + \frac{2 \zeta C - \frac{\delta B_0}{a}s^2}{2 s \mu^2}\arctan\left(\frac{x \zeta}{s}\right).
\end{equation}
Substituting to find $\Psi$,
\begin{equation}
\begin{aligned}
    \Psi &= \frac{\delta B_0 x^2}{2 a s} - \frac{\zeta x}{s}\left(\frac{\delta B_0}{2 a \zeta} x+ \frac{2 \zeta C - \frac{\delta B_0}{a}s^2}{2 s \mu^2}\arctan\left(\frac{x \zeta}{s}\right)\right)\\
    &= - x\left(\frac{C}{s^2} - \frac{\delta B_0}{2\zeta a}\right)\arctan\left(\frac{x\zeta}{s}\right).
\end{aligned}
\end{equation}

The boundary condition $\partial_x\Phi(x=a)=0$ sets the constant
    $C = \frac{\zeta \delta B_0 a}{2}$,
so 
\begin{equation}
    \Psi = x \frac{\delta B_0}{2 \zeta a}\left(1 - \frac{\zeta^2 a^2}{s^2}\right)\arctan\left(\frac{\zeta x}{s}\right).
\end{equation}
Inverse Laplace transforming in time, finally
\begin{align}
    \psi_1(x,t) &= \frac{B_0 \delta}{2 a \zeta}\bigg(-\frac{x}{t}\sin\left(\zeta x t\right)\\
    &+\zeta a^2\left(\cos(\zeta x t) -1\right) + \zeta^2 x t a^2 \text{Si}(\zeta x t)\bigg).
    \label{eq:ideal-hkt}
\end{align}
The sine integral Si is exactly the self-similar current sheet discussed previously. 

Inserting this growing current sheet as a source in the Voigt induction equation, we estimate
\begin{equation}
    \partial_t \psi_1 \approx \alpha_2 \partial_t J_{ideal} \sim \alpha_2 t.
\end{equation}
Therefore $\psi_1 \sim \alpha_2 t^2$ at early times. Here we note that the early behavior of the ideal solution \eqref{eq:ideal-hkt} is not dominated by the self-similar current sheet; this feature was not captured in the original treatment by Hahm and Kulsrud, who did not solve the ideal initial-value problem because the initial, transient behavior does not give rise to a singular current.

\begin{figure}
    \includegraphics[width = \columnwidth]{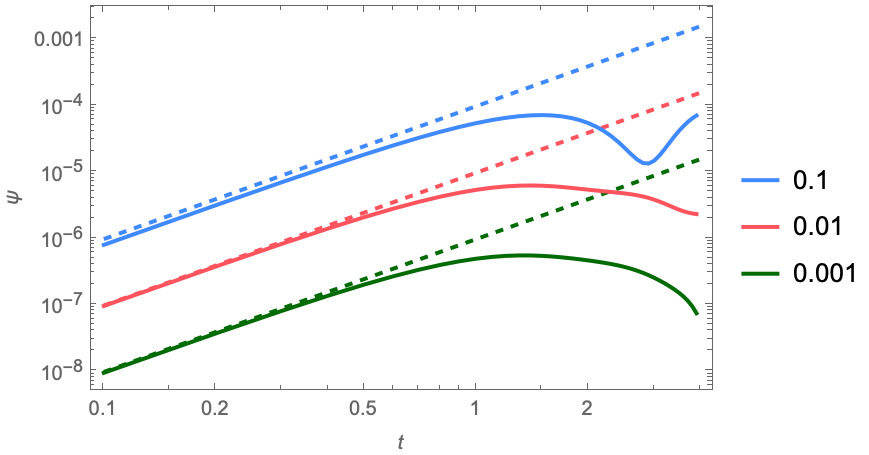}
    \caption{The early linear reconnection phase in the Voigt-HKT problem for $\alpha_1 = \alpha_2 = 0.1, 0.01, 0.001$. Dashed lines show $\psi_1 = \delta  \alpha_2t^2$. Note the rapid deviation from linear, ideal behavior.}
    \label{fig:lin-rx}
\end{figure}

Figure \ref{fig:lin-rx} verifies our prediction of the early linear tearing behavior, but the linear behavior is still not completely described by this ansatz. In the limit of small $\alpha$, the Voigt-regularized initial-value problem does admit an asymptotic description, as we now show.

In a thin layer about $x=0$ - so we drop $O(x^2)$ terms - the Fourier transform of equation \ref{eq:laplace} yields the ODE
\begin{equation}
    (1+\alpha_2 \xi^2) \Psi = \frac{\zeta^2}{s^2} \frac{d}{d\xi}\left(\frac{1}{\xi^2(1 + \alpha_1 \xi^2)}\frac{d}{d\xi}\left(\xi^2 \Psi \right)\right).
    \label{eq:lin_ode}
\end{equation}
In the limit $\alpha_1\xi^2\sim\alpha_2\xi^2 \gg 1$, the decaying solution at $\xi\to \infty$ is
\begin{equation}
    \Psi(\xi,s) = \frac{2C_0}{\Gamma(5/6)}\left(\frac{s}{6}\frac{\alpha_1^{1/2} \alpha_2^{1/2}}{ \zeta}\right)^{5/6}\xi^{1/2} K_{5/6}\left(\frac{s \alpha_1^{1/2}\alpha_2^{1/2}}{3 \zeta}\xi^3\right),
\end{equation}
where $K_{5/6}$ is a modified Bessel function of the second kind, and the high-$\xi$ behavior must match to the quasistationary solution at small $s$,
\begin{equation}
    \Psi_{qs} = \frac{C_0}{\xi^2} + C_1 \left(\frac{\xi}{3} + \alpha_1\frac{\xi^3}{5}\right),
\end{equation}
where $C_0(s) \sim s^{-1}$ is the outer-layer kink drive in $\psi$. Then the current scales as 

\begin{equation}
\begin{aligned}
    J(x=0,s) &\sim -\int_0^{\infty}\xi^2 \Psi(\xi,s) d\xi\\
    &\sim  \zeta^{1/3}s^{-4/3}(\alpha_1\alpha_2)^{-1/6}\int_0^{\infty}y^{1/6}K_{5/6}(y) dy
\end{aligned}
\end{equation}
in terms of an auxiliary variable $y = \frac{\sqrt{\alpha_1\alpha_2}}{3\zeta}s\xi^3$. Inverse Laplace transforming, the reconnected flux
\begin{equation}
    \psi_X(t) \sim \alpha_2 J(0,t) \sim \alpha_2^{5/6}\alpha_1^{-1/6}t^{1/3}.
\end{equation}

\begin{figure}
    \centering
    \begin{subfigure}[a]{\columnwidth}
        \includegraphics[width = \columnwidth]{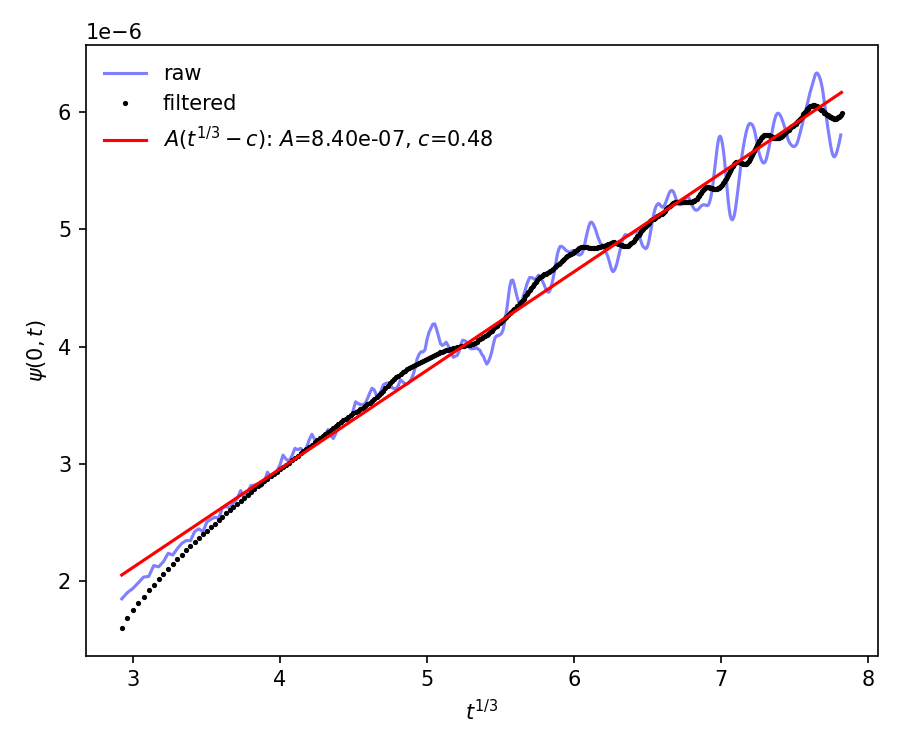}
        \caption{$\alpha_1 = \alpha_2 = 10^{-3}$, $\delta = 10^{-3}$, and $\mu = 0.05$.}
        \label{a_1e3_mu_5e2}
    \end{subfigure}
    \begin{subfigure}[b]{\columnwidth}
        \includegraphics[width = \columnwidth]{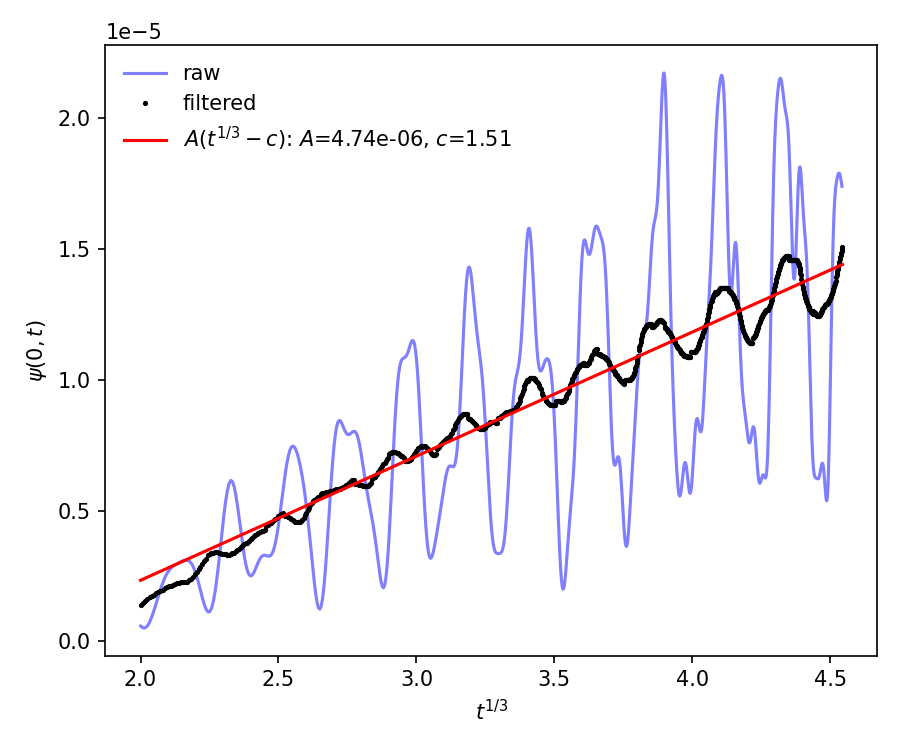}
        \caption{$\alpha_1 = \alpha_2 = 10^{-2}$, $\delta = 10^{-3}$, and $\mu = 0$.}
        \label{a_1e2_mu_0}
    \end{subfigure}
    \caption{Reconnected flux in the linear regime fitted with $\psi \sim t^{1/3}$ after filtering out fluctuations.}
    \label{fig:t13}
\end{figure}

The factor of $\alpha_2$ in the induction equation makes the momentum and inductive regularization parameters contribute asymmetrically to reconnection. Larger $\alpha_2$ accelerates reconnection, while $\alpha_1$ weakly decelerates it. The $t^{1/3}$ reconnection rate is verified in figure \ref{fig:t13}, including a case with $\mu=0.05$ to damp out Alfv\'enic oscillations that obscure the growth trend.

\begin{figure}
    \centering
    \includegraphics[width = \columnwidth]{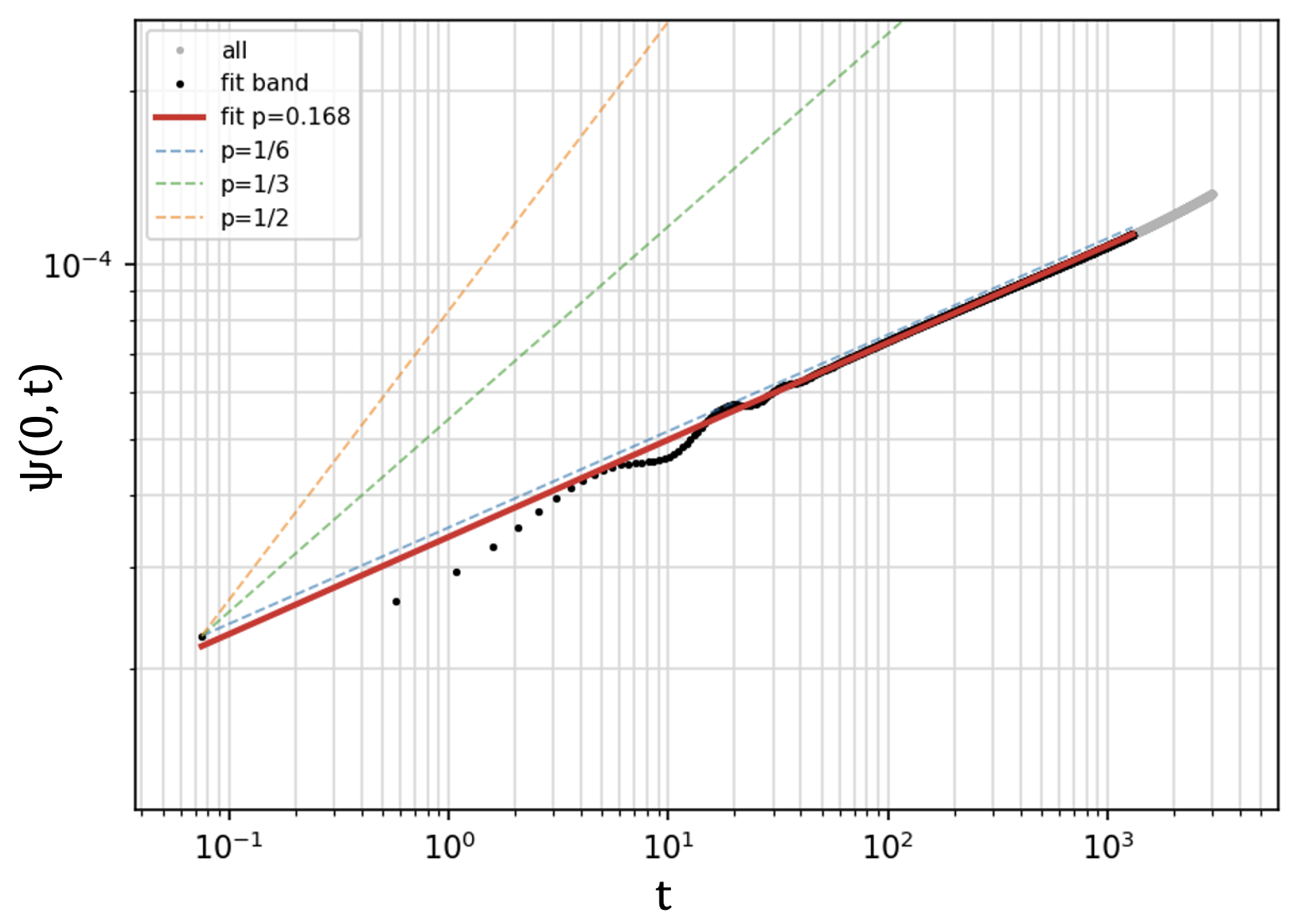}
    \caption{Reconnected flux in the linear regime fitted with $\psi \sim t^{1/6}$ with parameters $\eta = 10^{-6}, \nu = 3 \times 10^{-4}, \mu = 8\times 10^{-2}, \alpha_1 = 10^{-4}, \alpha_2 = 10^{-2}$, and $\delta = 3\times 10^{-3}$.}
    \label{fig:p16}
\end{figure}

For completeness, the generalization to finite dissipation follows by observing that, after Laplace transforming, $\eta$ and $\nu$ are equivalent to $s$-dependent $\alpha_1$ and $\alpha_2$, i.e., $\alpha_{2,\text{eff}} = \alpha_2 + \eta/s$ and $\alpha_{1,\text{eff}} = \alpha_1 + \nu/s$. Then 
\begin{equation}
    \psi_X(s) \sim \zeta^{1/3} \frac{(\alpha_2 s + \eta)^{5/6}}{s^2 (\alpha_1 s + \nu)^{1/6}}.
    \label{eq:gen-lin}
\end{equation}
When $\alpha_1$ and $\alpha_2$ vanish, we recover standard visco-resistive scaling \citep{comisso2015} $\psi(t) \sim t$. If we decouple $\alpha_1$ and $\alpha_2$, equation \eqref{eq:gen-lin} also yields a power law $\psi(t) \sim (t-t_0)^{p}$ when $\alpha_1=0$ and $\eta =0$, so $\psi(t)\sim t^{1/6}$; or when $\alpha_2=0$ and $\nu = 0$, so $\psi(t) \sim t^{7/6}$. The former case is verified in figure \ref{fig:p16}. These limiting cases represent distinct dominant balances in the governing ODE \eqref{eq:lin_ode}, including viscosity and resistivity. When regularization and dissipation effects are all non-negligible, equation \eqref{eq:gen-lin} no longer yields a clean power law after inverse transformation back to the time domain. Flow damping due to nonzero $\mu$ is not captured in the thin-layer, linear analysis.

\color{black}

\section{Nonlinear theory}
\label{sec:nonlin}
In this section we describe three qualitatively distinct regimes for the island width dynamics of the Voigt-HKT problem. Two of these regimes are distinguished by the ratio $\delta_{VR}/w$, where the visco-resistive length scale \citep{fitzpatrick2003, porcelli1987} is $\delta_{VR} = (\eta \nu)^{1/6} (k B')^{-1/3}$. The regularization length scales do not appreciably affect the nonlinear island dynamics until $\alpha_{1,2}^{1/2}\sim w_{sat}$, i.e., the Voigt scale becomes comparable to the island size. We note that Voigt-dominated reconnection foregoes the formation of a nonlinear, ''Waelbroeck'' current sheet \citep{wang1992}.

\begin{figure}
    \includegraphics[width = \columnwidth]{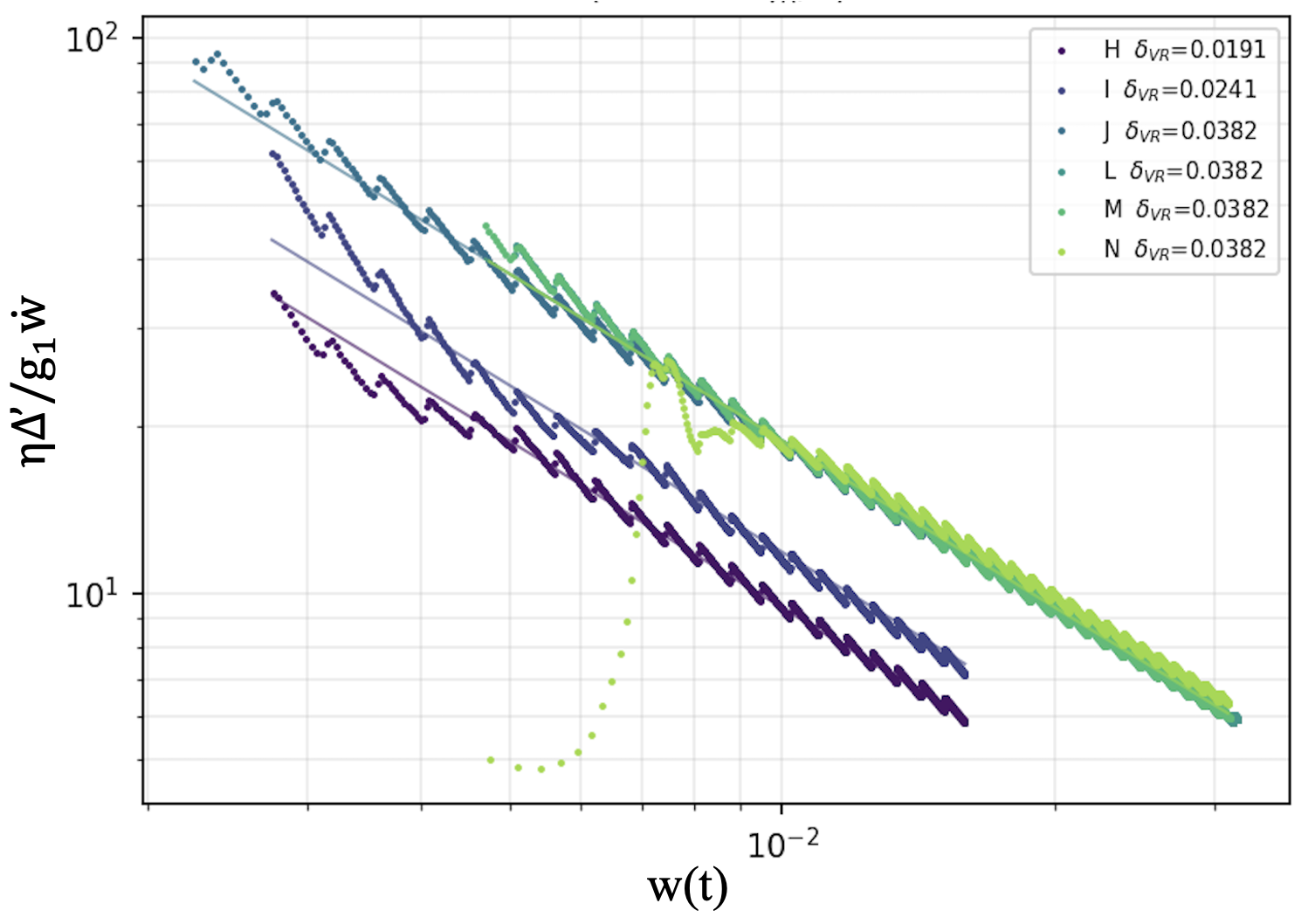}
    \caption{Viscous braking slows island growth compared to the Rutherford model by a factor $\sim w/\delta_{VR}$, independent of $\mu, \alpha_1$, and $\alpha_2$. Plotted runs all have $L_y = 3.5$. Other parameter values may be found in table \ref{fig:param-block}. Solid lines are $C \delta_{VR}/w$.}
    \label{fig:dvr_over_w}
\end{figure}

When the island width is small ($\delta_{VR} \gtrsim w$), viscosity slows island growth in accordance with the linear reconnection rate described previously \citep{comisso2015},
\begin{equation}
    \frac{g_1}{\eta}\frac{d w}{dt} = \Delta' \frac{w}{C \delta_{VR}}
\end{equation}
where $C\simeq 4.9$ is a dimensionless constant fitted across the six cases plotted in figure \ref{fig:dvr_over_w}. The theoretical value from Comisso's visco-resistive dispersion is $C \simeq 5.11$, in rough agreement with our numerical results. When the island width is small compared to the visco-resistive length scale, viscous slowing persists to saturation.

\begin{table}
\resizebox{\columnwidth}{!}{%
\begin{tabular}{|l|L|L|L|L|L|L|}
\hline
 Case & \eta    & \mu     & \nu     & \alpha_1 & \alpha_2 & \delta  \\
A & 3.6 \times 10^{-6}  & 10^{-7}            & 3.6 \times 10^{-6} & 10^{-3}          & 10^{-4} & 3 \times 10^{-2}  \\
B & 1.33 \times 10^{-6} & 10^{-7}            & 9.8 \times 10^{-6} & 10^{-3}          & 10^{-4} & 2 \times 10^{-2}  \\
F & 10^{-5}             & 7.5 \times 10^{-3} & 5 \times 10^{-7}   & 4 \times 10^{-3} & 10^{-4} & 6 \times 10^{-2}  \\
H & 10^{-5}             & 10^{-3}            & 1.56\times 10^{-5} & 10^{-4}          & 10^{-6} & 2.5\times 10^{-4} \\
I & 4\times 10^{-5}     & 10^{-3}            & 1.56\times 10^{-5} & 10^{-4}          & 10^{-6} & 2.5\times 10^{-4} \\
J & 10^{-5}             & 10^{-3}            & 10^{-3}            & 10^{-4}          & 10^{-6} & 2.5\times 10^{-4} \\
L & 10^{-5} & 10^{-3} & 10^{-3} & 10^{-4}  & 10^{-6}  & 10^{-3} \\
M & 10^{-5} & 10^{-3} & 10^{-3} & 10^{-3}  & 10^{-6}  & 10^{-3} \\
N & 10^{-5} & 10^{-3} & 10^{-3} & 10^{-6}  & 10^{-3}  & 10^{-3} \\
\hline
\end{tabular}
}
\caption{Parameter values for the cases used in figures \ref{fig:dvr_over_w} and \ref{fig:ruth_comp}.}
\label{fig:param-block}
\end{table}

\begin{figure}
    \includegraphics[width = \columnwidth]{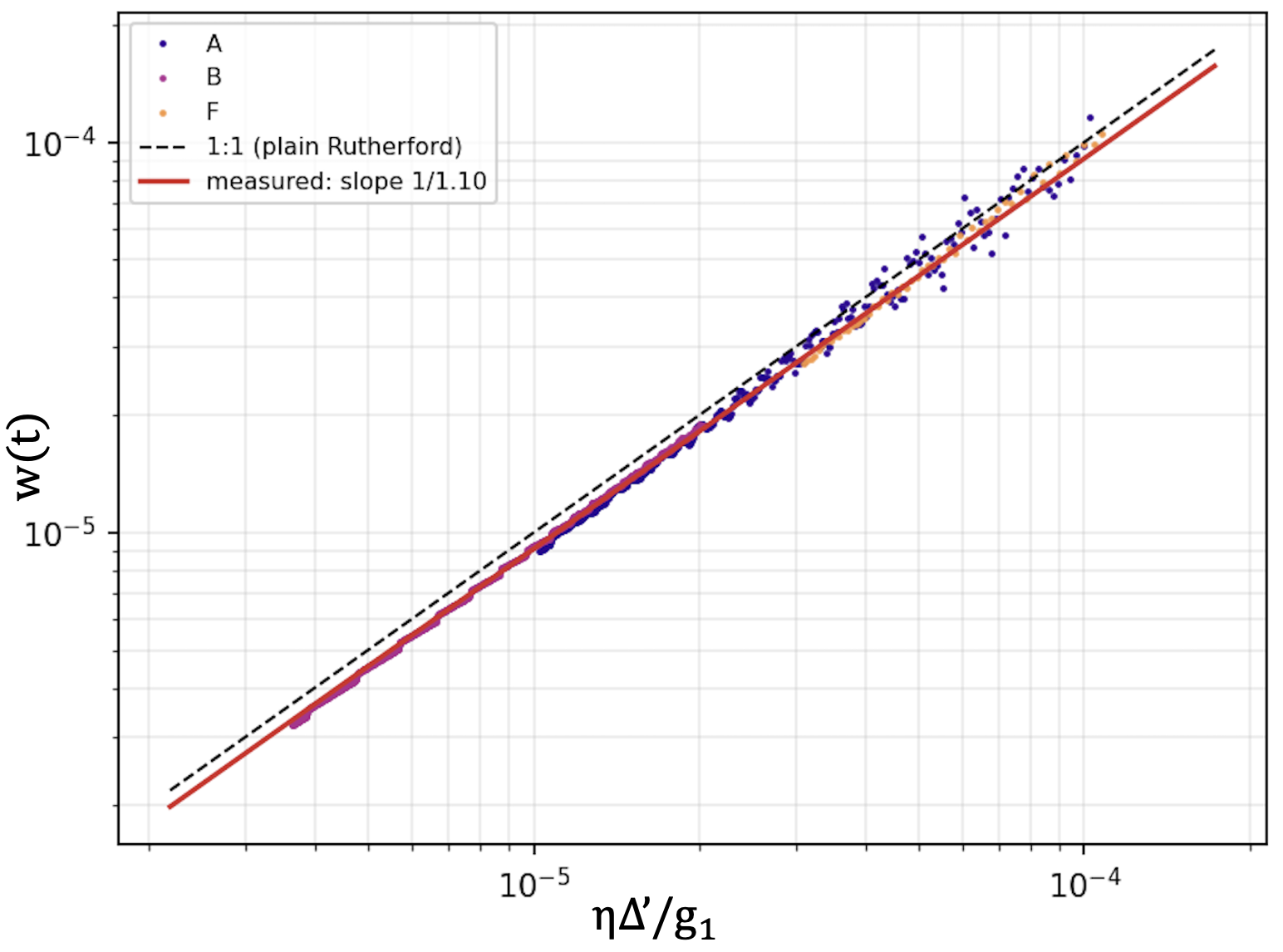}
    \caption{Comparison of $w(t)$ to the Rutherford model for cases A, B, and F, all with $L_y =3.5$. Other parameter values may be found in table \ref{fig:param-block}.}
    \label{fig:ruth_comp}
\end{figure}

When $w \gtrsim 3 \delta_{VR}$, the classical Rutherford model \citep{rutherford1973}
\begin{equation}
    \frac{g_1}{\eta}\frac{d w}{dt} = \Delta'(t)
    \label{eq:ruth}
\end{equation}
accurately describes the dynamics. Here the time-dependent tearing parameter
\begin{equation}
\begin{aligned}
    \Delta'(t) &= \frac{2k}{\sinh(kL_x/4)}\left(\frac{\delta L_x^2}{8 a\psi_s(t)} - \cosh(kL_x/4)\right)\\
    &= \frac{4k \delta a^2}{\sinh (kL_x/4)} \left(\frac{1}{w(t)^2} - \frac{1}{w_{sat}^2}\right),
    \label{eq:delta-prime}
\end{aligned}
\end{equation}
and the geometric coefficient 
\begin{equation}
    g_1 = \frac{1}{\sqrt{2} \pi} \int_{-1}^{1} \frac{\left( \int_{0}^{\theta_{max}} \frac{\cos \theta d\theta}{\sqrt{\Omega + \cos \theta}} \right)^2}{\int_{0}^{\theta_{max}} \frac{d\theta}{\sqrt{\Omega + \cos \theta}}} d\Omega \simeq 0.823.
\end{equation}
Figure \ref{fig:ruth_comp} shows agreement between numerical $w(t)$ and equation \ref{eq:ruth} to within a few percent.

\begin{figure}
    \includegraphics[width = \columnwidth]{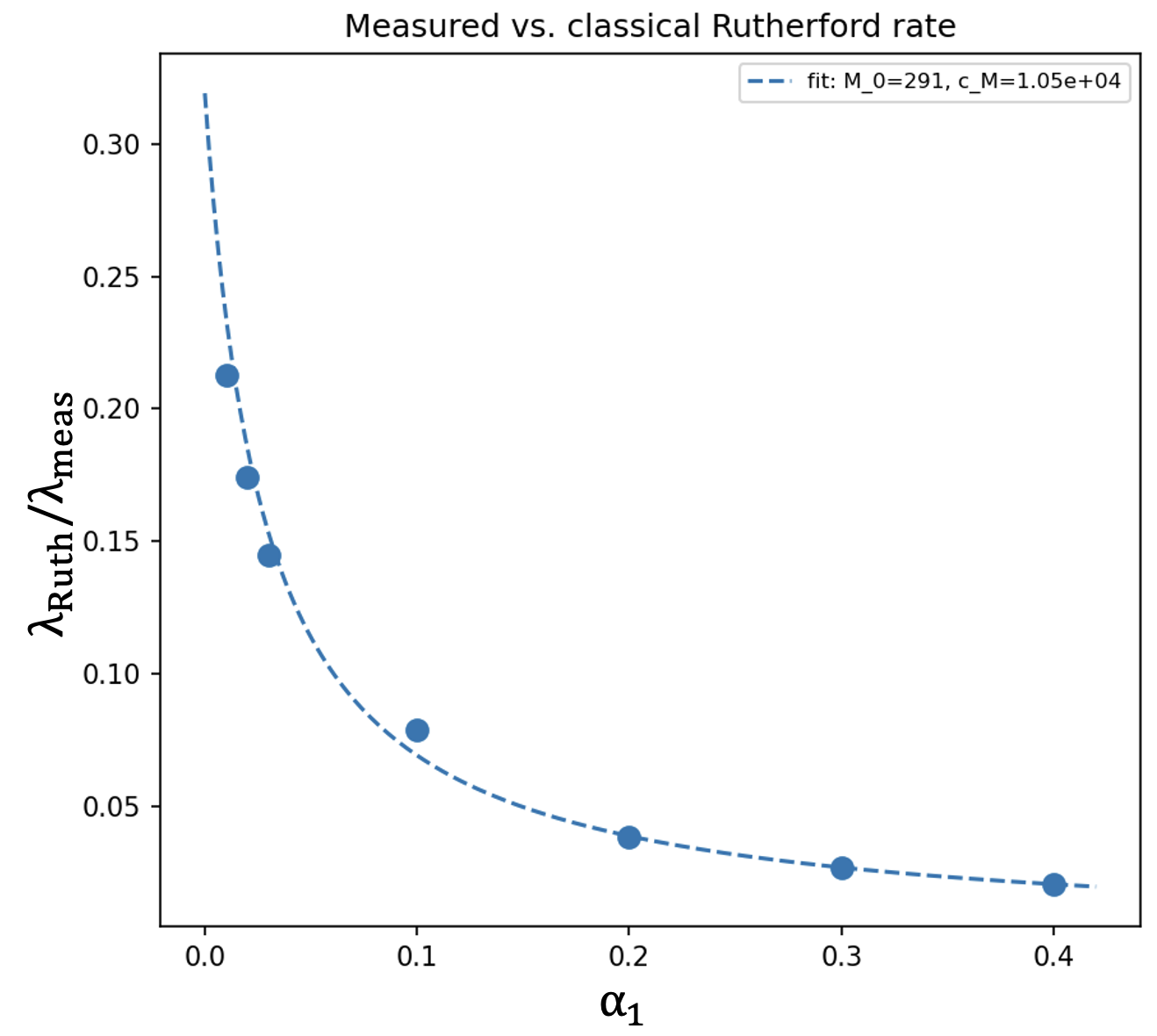}
    \caption{The exponential saturation rate $\lambda_{meas}$ compared to the linearized Rutherford value $\lambda_{Ruth}$, fit to $(M_0 + c_M \alpha_1)^{-1}$ for $\delta = 0.1, L_y = 7.0, L_x = 1.0, \nu = \eta = 10^{-4}$, and $\mu = 10^{-3}$. Note that $c_M/M_0 `\simeq 36.1$ is close to the naive prediction $k^2 = (2\pi)^2 \simeq 39.5.$}
    \label{fig:lam_comp}
\end{figure}

Finally, linearization of equation \eqref{eq:ruth} about the saturated island width $w_{sat}$ yields an exponential relaxation of $\tilde{w}\equiv w(t) -w_{sat}$,
\begin{equation}
    \frac{d\tilde{w}}{dt}\simeq-\frac{8 k \eta \delta a^2}{g_1\sinh(k L_x/4) w_{sat}^2}\tilde{w}.
    \label{eq:ruth-exp}
\end{equation}
Linear modes have all frequencies reduced by a $k$-dependent factor $\sim(1+\alpha_{1,2} k^2)$. Therefore, a naive scaling estimate is that the linear decay rate of equation \eqref{eq:ruth-exp} will be reduced by that factor, assuming that the outer-scale $k = \frac{L_y}{2\pi}$ is the relevant mode number. Fitting the late-time island saturation rate across a range of $\alpha_1=\alpha_2$ gives excellent agreement with numerics, as shown in figure \ref{fig:lam_comp}.

\begin{figure}
    \includegraphics[width = \columnwidth]{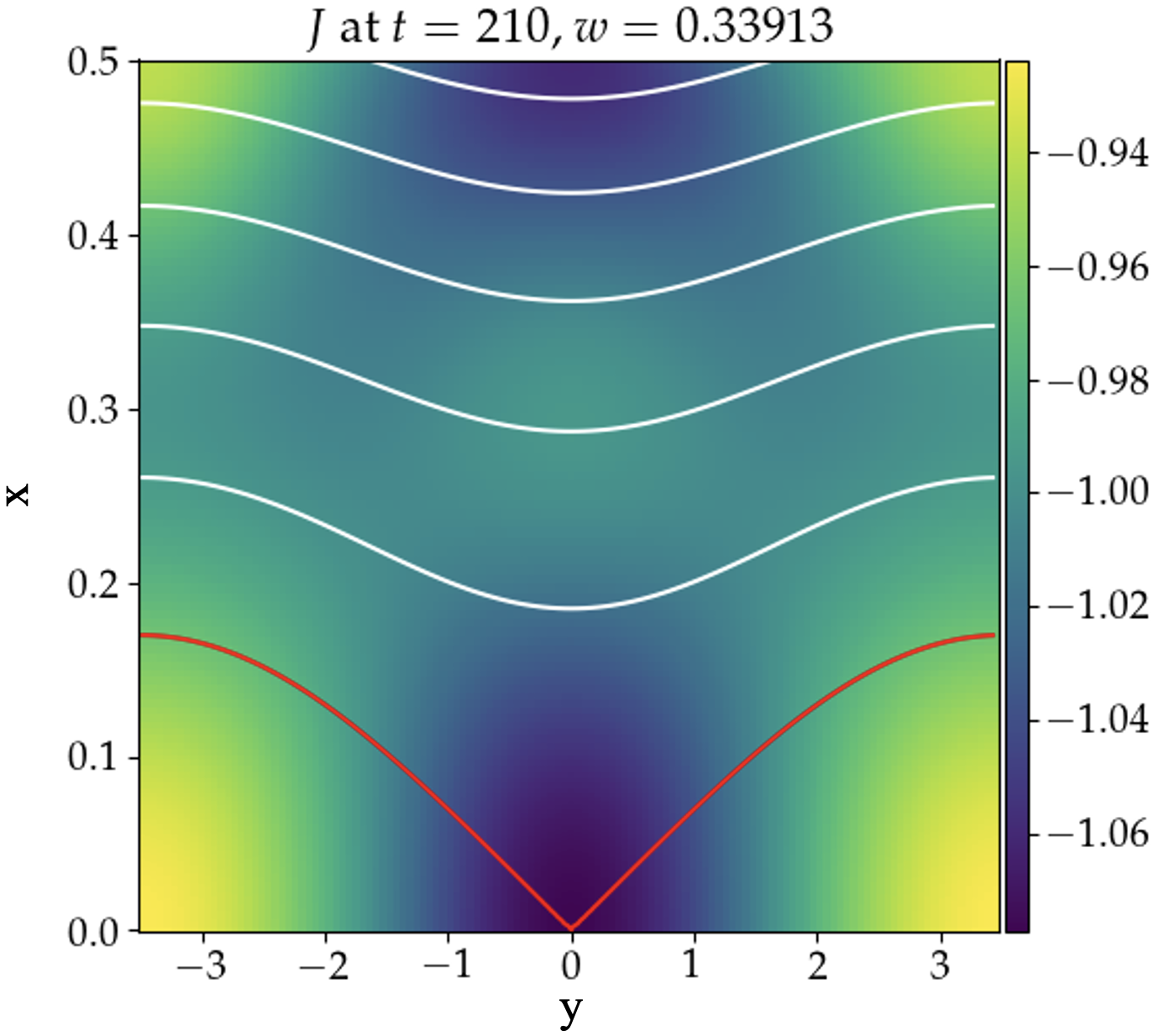}
    \caption{Current density at $t=210$ with $\mu = \nu = 10^{-3}$, $\eta = 10^{-4}$, $\delta = 0.1$ for $\alpha_1 = \alpha_2 = 0.5$ with timestep $0.05$. Note the variation in $J$ across the island.}
    \label{fig:j-psi}
\end{figure}

\begin{figure}
    \includegraphics[width = \columnwidth]{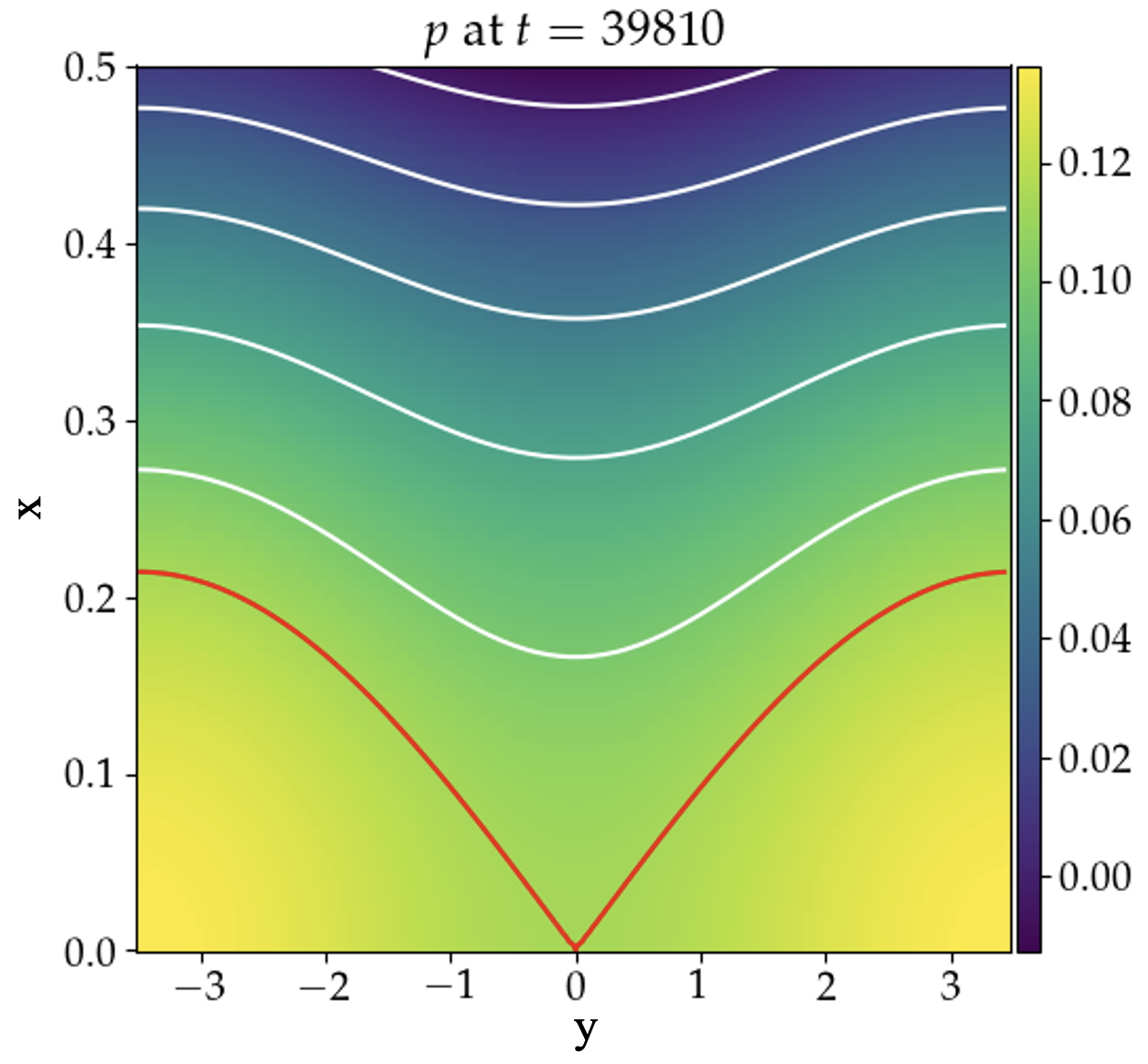}
    \caption{The saturated pressure in the case from figure \ref{fig:j-psi}. The final island width $w=0.42787$ is the same as that found in all other cases with $\delta=0.1$.}
    \label{fig:equil-p}
\end{figure}

\begin{figure}
    \includegraphics[width = \columnwidth]{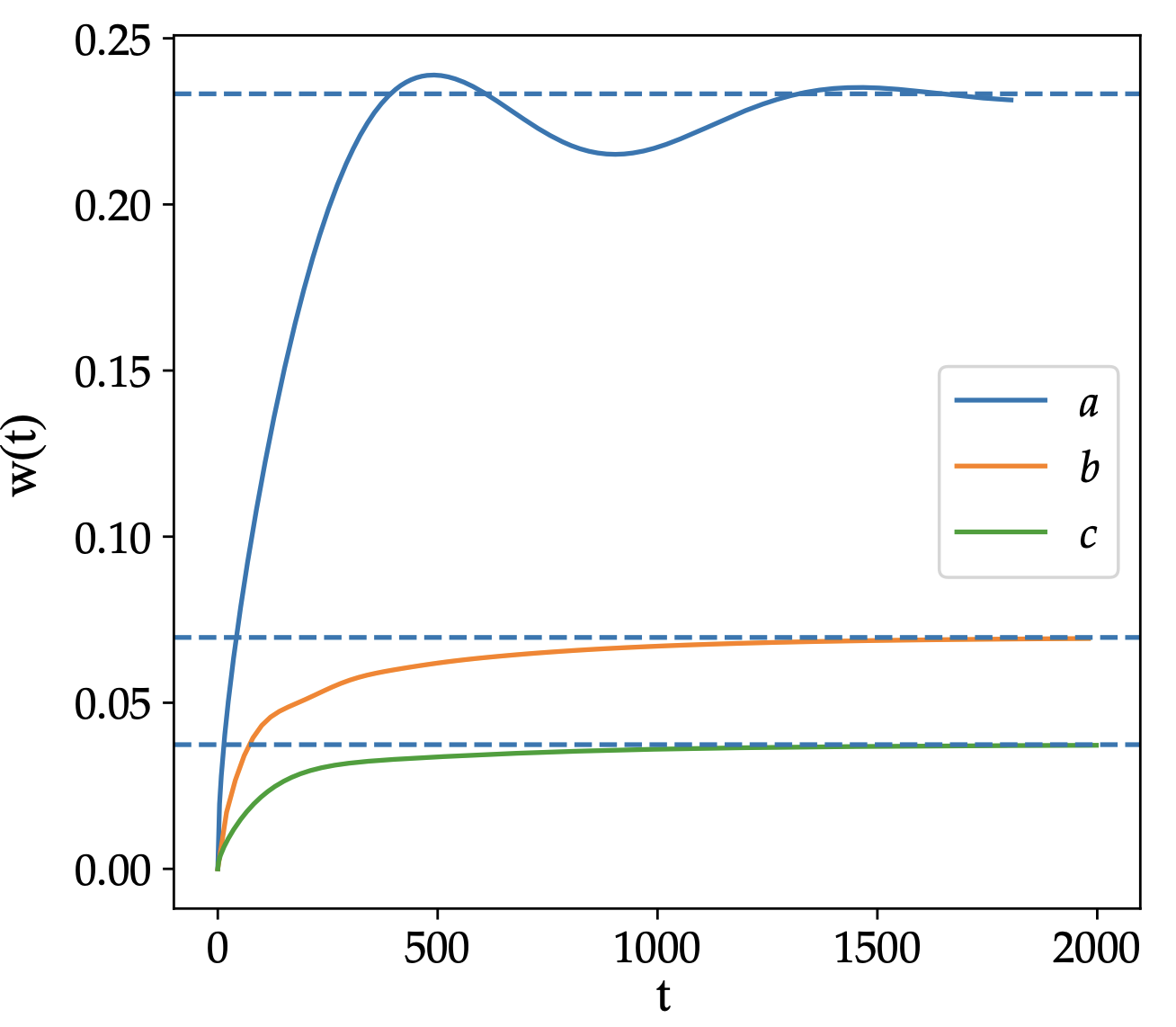}
    \caption{Island width evolution for cases (a) $\delta=0.03, L_y = 7$; (b) $\delta = 0.01, L_y = 1.5$; and (c) $\delta = 10^{-3}, L_y = 3.5$.}
    \label{fig:w-scan}
\end{figure}

Preliminary investigation of reconnection in Voigt-MHD found that, when resistivity is included, the saturated state was independent of the regularization and dissipation parameters \citep{huang2025}. The results presented here are consistent with this observation: Viscosity and Voigt regularization do, in different parameter regimes, slow island growth, but they do not affect the reconnection drive $\Delta'$, which ultimately determines saturation. Figure \ref{fig:equil-p} is one of many examples; for all parameter values tested with $\delta=0.1$, the same saturated island width is found. Moreover, if the final state is independent of the regularization parameters, the island width must be that predicted by ideal theory. The outer layer first described by Hahm and Kulsrud has an island of width
\begin{equation}
    w = \frac{L_x}{2}\sqrt{\frac{\delta}{\cosh(kL_x/2)}}.
    \label{eq:w-ideal}
\end{equation}
Figure \ref{fig:w-scan} demonstrates excellent agreement between the ideal prediction and the saturated Voigt island widths. The island growth and saturation for $\delta=10^{-2}$ and $10^{-3}$ appears to be monotonic, as predicted by a Rutherford-like model. However, for $\delta=0.03$ the island width evolves as an underdamped oscillator, which cannot be modeled by a first-order, linear ODE. Moreover, the Rutherford model implicitly relies on $J = J(\psi)$, which is manifestly false at large $\alpha_2$, as shown in figure \ref{fig:j-psi}. Nevertheless, the large-$\alpha$ brake described above still captures the envelope decay and island saturation rate.

\section{Equilibrium}
\label{sec:equil}
In this section we consider the long-time behavior of the Voigt system with dissipation and present evidence that the theorem of Constantin and Pasqualotto \citep{constantinp2023} which guarantees convergence to an ideal equilibrium may be generalized to this case. The system \eqref{eq:momentum}--\eqref{eq:constraints} conserves a modified energy\citep{huang2025} (which may be identified with the usual norm on the Sobolev space $H^1$)

\begin{align}
    \frac{1}{2}\frac{d}{dt}\int & d^3 x \left(B^2 + \alpha_2 J^2 + u^2 + \alpha_1 \omega^2 \right) = \\
    &-\int d^{3}x\left(\nu\omega^{2}+\eta J^{2}+\mu u^2\right)-\int d^{3}x\boldsymbol{E}_{\text{ext}}\cdot\boldsymbol{J},\label{eq:energy}
\end{align}
where the vorticity $\boldsymbol{\omega} = \nabla \times \boldsymbol{u}$. In particular, when the applied field vanishes, this energy is monotonically decaying in time. Nonzero $\alpha_2$ grants control on the $L_2$ norm of $J$, prohibiting the formation of $\delta$-function current sheets. The additional friction term adds a sign-definite dissipation on the RHS of equation \eqref{eq:energy}, so we do not anticipate worsened regularity from $\mu\neq0$. In steady state with nonzero $\eta$, equation (\ref{eq:induction}) gives a residual flow
\begin{equation}
    \boldsymbol{u}\times\boldsymbol{B}=\eta \boldsymbol{J} + \boldsymbol{E}_{ext}.
\end{equation}

When $\mu$ vanishes, steady-state momentum balance reads
\begin{equation}
    \boldsymbol{J}\times \boldsymbol{B} = \nabla p -\nu \nabla^2 \boldsymbol{u}+\boldsymbol{u}\cdot\nabla \boldsymbol{u}.
\end{equation}
The deviation from ideal MHS equilibrium $\boldsymbol{J}\times\boldsymbol{B}=\nabla p$ is an undesirable complication in the context of 3D equilibrium computation. On the other hand, the theorems of Constantin and Pasqualotto \citep{constantinp2023,pasqualottoThesis} guarantee (when $\mu = \eta = 0$, $\boldsymbol{E}_{ext}=0$, and for sufficiently regular initial conditions) an asymptotic state of the system \eqref{eq:momentum}--\eqref{eq:constraints} that exactly satisfies the ideal MHS equilibrium relation. We conjecture, with numerical evidence, that the inclusion of nonzero $\mu\text{, }\eta$ and $\boldsymbol{E}_{ext}$ will nonetheless yield a smooth asymptotic steady state.

\begin{figure}
    \includegraphics[width = \columnwidth]{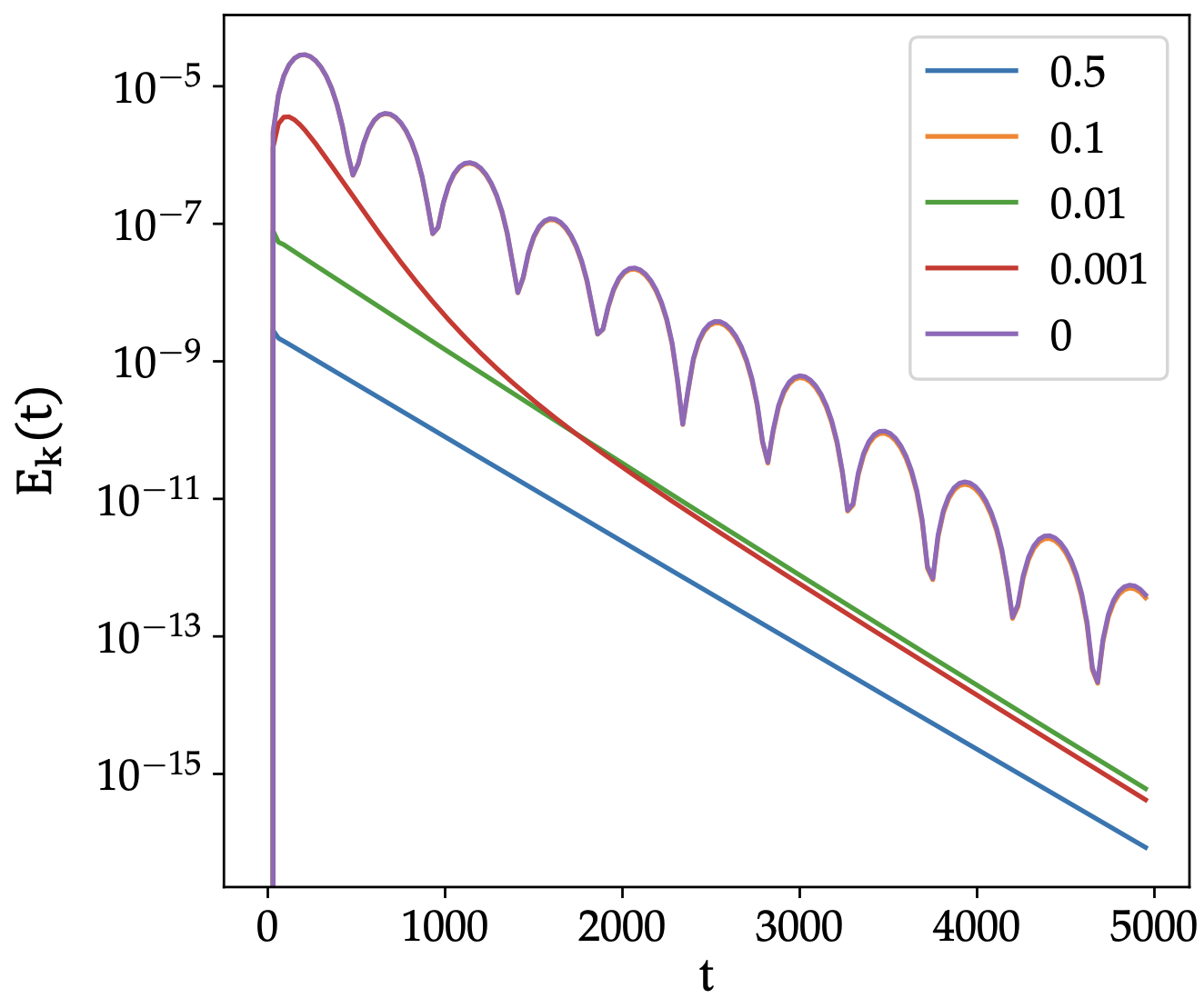}
    \caption{Kinetic energy decay for $\delta = 0.1$ and $\alpha_1=\alpha_2=0.5$ with $\mu =\nu= 0.5, 0.1, 0.01$ and $10^{-3} $ and $\mu=0$, $\nu=10^{-3}$ with timestep 30.}
    \label{fig:Ek-mu-scan}
\end{figure}

\begin{figure}
    \includegraphics[width = \columnwidth]{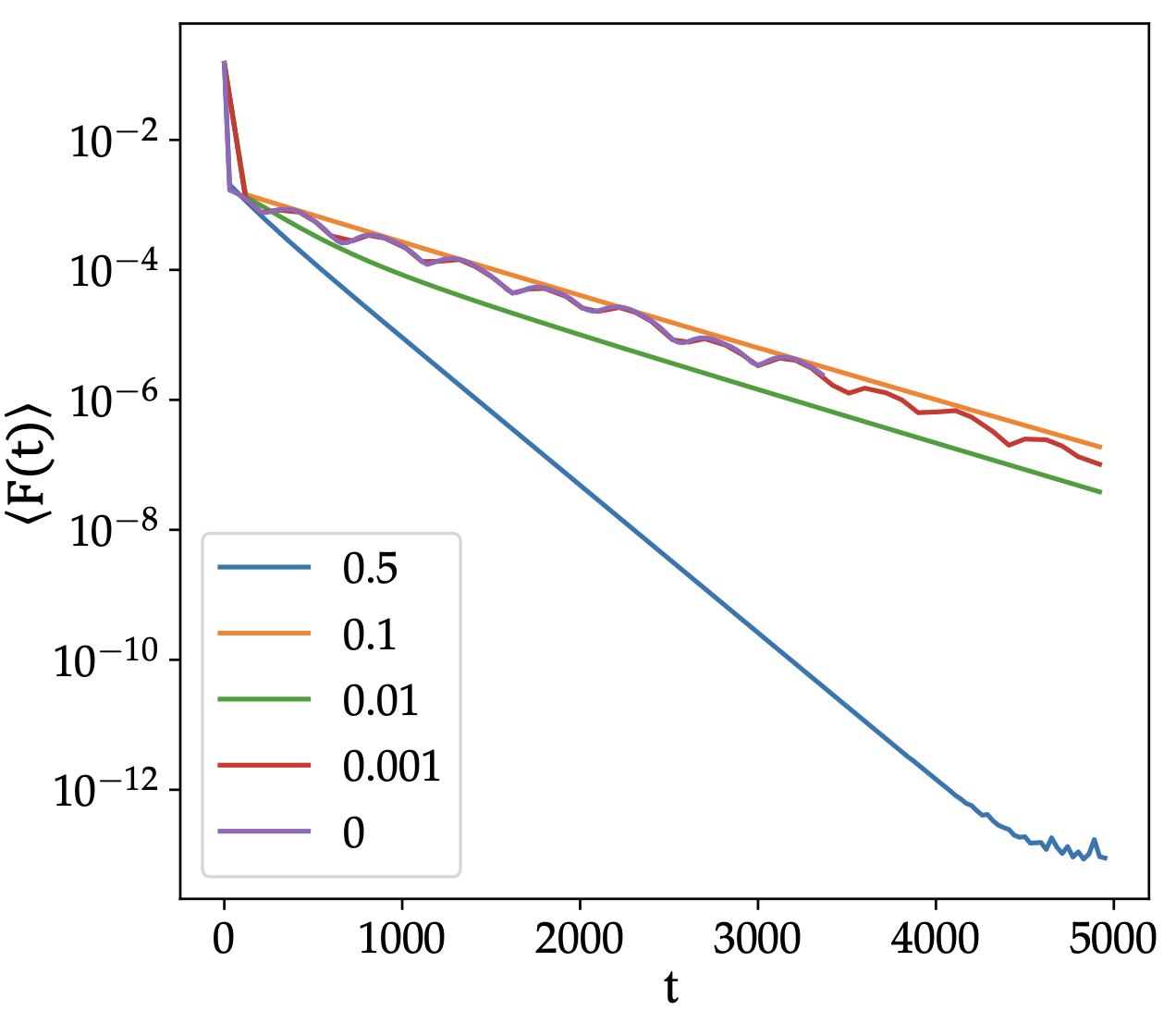}
    \caption{Volume-averaged force residual decay for $\delta = 0.1$ and $\alpha_1=\alpha_2=0.5$ with $\mu =\nu= 0.5, 0.1, 0.01$ and $10^{-3} $ and $\mu=0$, $\nu=10^{-3}$ with timestep 30.}
    \label{fig:residual-mu-scan}
\end{figure}

In figure \ref{fig:Ek-mu-scan}, the strong decay of the kinetic energy $\int u^2$ induced by large values of $\nu$ and $\mu$ is evident. Figure \ref{fig:residual-mu-scan} shows the volume-averaged force residual $\vert \boldsymbol{J}\times\boldsymbol{B}-\nabla p\vert$, which rapidly decays to machine-precision zero in the case $\mu=0.5$, indicating that the ideal MHS equilibrium condition is satisfied, and that there is some speed-up in convergence compared to cases without friction. We also note that the form of the applied field implies that, when the flow has entirely dissipated, the final current must equal the initial current. Therefore, the addition of friction restricts the final equilibrium to differ from the initial state by only a vacuum magnetic field. 

The numerical results presented in this study suggest that the Voigt-regularized MHD equations behave as a strongly damped system where the magnetic topology is constrained by a competition between the inductive scale $\alpha$ and the dissipative scales $\eta$ and $\nu$. While the Constantin-Pasqualotto theorems provide a rigorous foundation for the existence and uniqueness of solutions in 2D MHD, our findings indicate that the inclusion of the Voigt-brake effect alters the nonlinear saturation manifold.

\section{Conclusion}
Our investigation of Voigt-regularized MHD has uncovered several key insights: In section \ref{sec:lin}, we saw that the linear phase of forced reconnection is rapid compared to resistive MHD, forgoing the need for a singular current sheet to develop. In section \ref{sec:nonlin}, we found that the island evolution has subtle deviations from a Rutherford-like model, but the saturated width is insensitive to the choice of regularization and dissipation parameters. Finally, in section \ref{sec:equil} we presented the first evidence that Voigt-regularized MHD with friction can generate precise MHS equilibria without flows - and, we reiterate, that the choice of regularization parameters does not meaningfully affect the final equilibrium state.

It remains an open question how these phenomena will be complicated by the effects of reactor-relevant geometry. 
The highly shaped, 3D magnetic fields of optimized stellarators will undoubtedly present many challenges, which we hope to address in future work.

\begin{acknowledgments}
The authors would like to thank W. Sengupta, S. Buller, M. Ruth, J. Burby, and the late R. Dewar for their helpful insights.

This research was supported by the Simons Foundation/SFARI (grant No. 560651, A.B.) and the Department of Energy SciDAC HifiStell grant (DE-SC0024548) (until March 31, 2025). 
\end{acknowledgments}

\appendix
\section{Voigt Helicity and Electron Inertia}
\label{app:helicity}
There is an intriguing similarity between the induction equation \eqref{eq:induction} and that of inertial MHD, as studied in Hamiltonian reconnection
\citep{grasso1999}. There one finds
\begin{equation}
    \partial_t \left(\boldsymbol{B}-d_e^2\nabla^2 \boldsymbol{B}\right) = \nabla \times \left(\boldsymbol{u}\times \left(\boldsymbol{B}-d_e^2\nabla^2 \boldsymbol{B}\right)\right).
\end{equation}
The electron skin depth $d_e$ plays a similar role to the Voigt regularization length $\alpha_2^{1/2}$. However, inertial MHD has a noncanonical Hamiltonian structure\citep{morrison1984} that is apparent in that $d_e$ appears on both sides of the induction equation, so $\boldsymbol{B}-d_e^2 \nabla^2 \boldsymbol{B}$ is dual to an advected one-form with associated conserved helicity
\begin{equation}
    \mathcal{K}_e = \int d^3 x \left(\boldsymbol{A}-d_e^2 \nabla^2 \boldsymbol{A}\right)\cdot \left(\boldsymbol{B}-d_e^2 \nabla^2 \boldsymbol{B}\right),
\end{equation}
which effectively counts linkages of a field $\tilde{\boldsymbol{B}}\equiv\boldsymbol{B}-d_e^2 \nabla^2 \boldsymbol{B}$. 
The Voigt MHD induction equation does not respect such a variational structure. Nevertheless, as proved by Constantin and Pasqualotto\citep{constantinp2023}, it does conserve a ``helicity''
\begin{equation}
    \mathcal{K} = \int d^3x \boldsymbol{B}\cdot \left(\boldsymbol{A} - \alpha_2 \nabla^2 \boldsymbol{A}\right),
\end{equation}
which we recognize as the weighted sum of the familiar magnetic helicity $\int \boldsymbol{A}\cdot\boldsymbol{B}$ and the current helicity $\int \boldsymbol{B}\cdot \boldsymbol{J}$. In other words, Voigt MHD allows change in magnetic topology, but only when the topology of current lines changes to compensate.

\bibliographystyle{apsrev4-1}
\bibliography{forced_rx}

\end{document}